\documentclass{llncs}

\usepackage{fullpage}
\usepackage{verbatim}
\usepackage{harvard}
\usepackage{fancybox}
\usepackage{graphicx}
\usepackage{amsmath}
\usepackage{wrapfig}
\usepackage{textcomp}
\usepackage{sidecap}
\usepackage[lined,commentsnumbered]{algorithm2e}
\usepackage{amssymb}
\usepackage{mdframed}
\usepackage[colorlinks=true]{hyperref}
\usepackage{fullpage}
\newcommand\qedsymbol{$\blacksquare$}

\usepackage{pslatex}
\usepackage{booktabs}
\usepackage{verbatim}
\usepackage{color}
\usepackage{amsfonts}
\usepackage{amsmath}
\usepackage{amssymb}

\usepackage{xcolor}

\usepackage{proof}
\usepackage{fancybox}
\usepackage{bussproofs}

\mathchardef\mhyphen="2D

\newcommand{\omegarule}{\ensuremath{\omega\mhyphen \mbox{rule}}}

\newcommand{\lsort}[1]{%
  \ensuremath{\mbox{\textsf{#1}}}}
\newcommand{\defsort}[2]{%
  \newcommand{#1}{\lsort{#2}}}
\defsort{\Action}{Action}
\defsort{\Time}{Time}
\defsort{\Self}{Self}
\defsort{\SortName}{SortName}

\defsort{\Agent}{Agent}
\defsort{\Entrant}{Entrant}
\defsort{\ActionType}{ActionType}
\defsort{\Moment}{Moment}
\defsort{\Boolean}{Boolean}
\defsort{\PayOut}{PayOut}
\defsort{\Fluent}{Fluent}
\defsort{\Event}{Event}
\defsort{\Object}{Object}
\defsort{\Numeric}{Numeric}

\newcommand{\lsymbol}[1]{%
  \ensuremath{\mathit{#1}}}
\newcommand{\defsymbol}[2]{%
  \newcommand{#1}{\lsymbol{#2}}}
\defsymbol{\action}{action}
\defsymbol{\initially}{initially}
\defsymbol{\holds}{holds}
\defsymbol{\mirrored}{mirrored}
\defsymbol{\happens}{happens}
\defsymbol{\clipped}{clipped}
\defsymbol{\initiates}{initiates}
\defsymbol{\terminates}{terminates}
\defsymbol{\prior}{prior}
\defsymbol{\interval}{interval}

\defsymbol{\does}{does}
\defsymbol{\plans}{plans}
\defsymbol{\act}{act}
\defsymbol{\react}{react}
\defsymbol{\payTot}{pay_{tot}}
\defsymbol{\fight}{fight}
\defsymbol{\coop}{coop}
\defsymbol{\enter}{enter}
\defsymbol{\stayout}{stayout}
\defsymbol{\learns}{learns}
\defsymbol{\redsplotched}{red\mhyphen splotched}
\defsymbol{\hassplotch}{has\mhyphen splotch}
\defsymbol{\payoff}{payoff}

\defsymbol{\capable}{capable}
\defsymbol{\destroyed}{destroyed}

\defsymbol{\removesplotch}{remove\mhyphen splotch}
\defsymbol{\remove}{remove}

\defsymbol{\cogito}{cogito}
\defsymbol{\named}{named}
\defsymbol{\deter}{deter}
\defsymbol{\enhance}{enhance}
\defsymbol{\attack}{attack}
\defsymbol{\cost}{cost}

\newcommand{\lconstant}[1]{%
  \ensuremath{\mbox{\textsf{#1}}}}
\newcommand{\defconstant}[2]{%
  \newcommand{#1}{\lconstant{#2}}}
\defconstant{\Enter}{Enter}
\defconstant{\StayOut}{StayOut}
\defconstant{\Fight}{Fight}
\defconstant{\Acquiesce}{Acquiesce}
\defconstant{\cs}{cs }
\defconstant{\estimate}{estimate}

\newcommand{\lmodality}[1]{%
  \ensuremath{\mathbf{#1}}}
\newcommand{\defmodality}[2]{%
  \newcommand{#1}{\lmodality{#2}}}
\defmodality{\common}{C}
\defmodality{\knows}{K}
\defmodality{\believes}{B}
\defmodality{\perceives}{P}
\defmodality{\mental}{M}
\defmodality{\ought}{O}

\defmodality{\desires}{D}
\defmodality{\intends}{I}
\defmodality{\says}{S}

\defsymbol{\wipe}{wipe\mhyphen fore\mhyphen head}

\defsymbol{\us}{\mathsf{us}}
\defsymbol{\iran}{\mathsf{iran}}
\defsymbol{\israel}{\mathsf{israel}}
\defsymbol{\russia}{\mathsf{russia}}
\defsymbol{\T}{\mathsf{T}}
\defsymbol{\ugv}{\mathsf{ugv}}
\defsymbol{\uav}{\mathsf{uav}}
\defsymbol{\carrying}{carrying}
\defsymbol{\firefight}{firefight}
\defconstant{\main}{main}
\defsymbol{\mission}{mission}
\defconstant{\silence}{silence}

\defconstant{\solda}{\ensuremath{\mathsf{soldier_A}}}
\defconstant{\soldb}{\ensuremath{\mathsf{soldier_B}}}
\defconstant{\sold}{\ensuremath{\mathsf{soldier}}}
\defconstant{\commander}{\ensuremath{\mathsf{commander}}}

\defconstant{\enemyterritory}{enemyterritory}
\defconstant{\baseb}{\ensuremath{\mathsf{base_b}}}
\defsymbol{\move}{move}
\defsymbol{\allowed}{\mathbf{allowed}}
\defconstant{\past}{past}

\let\originalleft\left
\let\originalright\right
\renewcommand{\left}{\!\mathopen{}\mathclose\bgroup\originalleft}
\renewcommand{\right}{\aftergroup\egroup\!\originalright}

\defconstant{\robot}{Robot}

\newcommand{\pa}{\ensuremath{\mathsf{PA}}}

\newcommand{\pafor}{\ensuremath{\mathsf{\langle\pa,\rho^\omega\rangle}}}

\newcommand{\yes}{\ensuremath{\mathsf{yes}}}
\newcommand{\no}{\ensuremath{\mathsf{no}}}
\newcommand{\loops}{\ensuremath{\mathsf{loops}}}

\begin{document}
\title{Proof Verification Can Be Hard! \thanks{This abstract was
    accepted for a presentation at
    Computability in Europe 2014: \href{{http://cie2014.inf.elte.hu/?Accepted_Papers}}{http://cie2014.inf.elte.hu/?Accepted\_Papers}.}}
\author{Naveen Sundar Govindarajulu$^{2}$ \& Selmer Bringsjord$^{1,2}$}
{\institute{ Department of Computer Science$^{1}$, Department of Cognitive Science$^{2}$
\\ Rensselaer Polytechnic Institute (RPI) 
\\ Troy NY 12180 USA \\ \email{govinn2@rpi.edu $\bullet$
  selmer@rpi.edu}}
\maketitle
\keywords{restricted \omegarule, not-semi-decidable, limits of proof verification}

The generally accepted wisdom in computational circles is that pure
proof verification is a solved problem and that the computationally
hard elements and fertile areas of study lie in proof
discovery.\footnote{Conjecture generation in our experience is also
  commonly regarded to be genuinely difficult.}  This wisdom
presumably does hold for conventional proof systems such as
first-order logic with a standard proof calculus such as natural
deduction or resolution.  But this folk belief breaks down when we
consider more user-friendly/powerful inference rules.  One such rule
is the \textbf{restricted \omegarule}, which is \emph{not even
  semi-decidable} when added to a standard proof calculus of a
\emph{nice} theory.\footnote{\emph{Nice} theories are consistent,
  decidable, and allow representations \cite{ebb.flum.thomas}.
  Roughly put, if a theory allows representations, it can prove facts
  about the primitive-recursive relations and functions.  (See Smith
  \citeyear*{intro_godel_theorems_smith}.) A formal system (a theory
  $\Gamma$ and a proof calculus $\rho$) is
  decidable/semi-decidable/not-semi-decidable if the decision problem
  $\Gamma\vdash_\rho\gamma$ is
  decidable/semi-decidable/not-semi-decidable.} While presumably not a
novel result, we feel that the hardness of proof verification is
under-appreciated in most communities that deal with proofs.  A
proof-sketch follows.

We set some context before we delve into the sketch.  The formal
machinery and conventions follow \cite{ebb.flum.thomas}.  We assume
the standard apparatus of first-order logic and that we are concerned
only with theories of arithmetic, and machine checking and discovery
of proofs of theorems of arithmetic.  A theory $\Gamma$ is said to be
\emph{negation-incomplete (incomplete)} iff there is at least one
$\phi$ such that $\Gamma\not \vdash\phi$ and $\Gamma\not
\vdash\lnot\phi$.  As readers will recall, G\"{o}del's first
incompleteness theorem states that any sufficiently strong theory of
arithmetic that has certain desirable attributes is incomplete.  Peano
Arithmetic (\pa) is one of the smallest incomplete theories that
covers all of standard arithmetic.  One way to surmount incompleteness
is to add more user-friendly (or mathematician-friendly) rules of
inference.
 
The \omegarule\ is one such rule of inference.
The \omegarule\ can be added in order to complete proof calculi;
specifically, the \omegarule\ renders \pa\ complete.  This infinitary
rule is of the following form:

\begin{footnotesize}
\begin{equation*}
\indent \indent \indent \infer[{ \omegarule}]{\forall x\
  \phi(x)}{\phi(\overline{0}),\phi(\overline{1}) \ldots,}
\end{equation*}
\end{footnotesize}

\vspace{-0.14in} The above rule has an infinite number of premises and
is clearly not suitable for implementation.  A \emph{restricted
  \omegarule} is a finite form of the rule which still keeps
$\mathsf{PA}$ complete.

Assume that we have machines operating over representations of
numerals and proofs.  Then if we have a machine $\mathsf{m}_{\phi}$,
which for all $n\in\mathbb{N}$ and the formula $\phi$ with one free
variable, it produces a proof of $\phi(\overline{n})$ from some set of
axioms $\Gamma$.  That is, $\mathsf{m}_\phi: \overline{n} \mapsto
\rho(\Gamma,\phi(\overline{n}))$.\footnote{An accessible reference for
  the \omegarule\ is \cite{baker1992use}.  All these results, except
  the main argument in the present abstract, are available in
  \cite{ebb.flum.thomas,franzen2004transfinite}.}

Given this, one form of the restricted \omegarule\
is as follows:
\begin{footnotesize}
\begin{equation*}
\indent \indent \indent \infer{\forall x\ \phi(x)}{\Gamma \ \ \
  \mathsf{m}_\phi}
\end{equation*}
\end{footnotesize}

\vspace{-0.14in}
 
Though the restricted \omegarule\ can (as just seen) be written down
in full, complete checking of the rule is beyond any machine
implementation, since in the general case, a proof verification system
that handles the rule would be able to check in all possible cases
whether the program supplied halts with the correct proof.  A simple
proof of this limit is given in the appendix.  We feel that this
limitative result demonstrates that proof representation and proof
verification in mathematics can be a fertile area of study involving a
rich interplay between expressibility and computational costs.

\appendix
\section*{Appendix: Proof}
\begin{small}
\textbf{Theorem}: \pafor\ is not-semi-decidable
\begin{proof}
  Let \pafor\ denote the formal system comprised of \pa\ with a
  standard proof calculus $\rho$ augmented with the restricted
  \omegarule.  Assume that we are only talking about Turing machines
  which output exactly one of $\{\yes,\no\}$ on all inputs, or else go
  on forever without halting, e.g., \loops.  The inputs are numerals
  which encode natural numbers.

\textbf{Given}: \pafor\ is negation-complete and all its theorems are
true on the standard model
$\langle\mathbb{N};0,\mathsf{S,+,1}\rangle$.

The following three statements can be coded up as arithmetic
statements in the language of \pa.

 \begin{quote}
\begin{enumerate}
\item Machine $m$ on input $n$ halts with \yes
\item Machine $m$ on input $n$ halts with \no
\item Machine $m$ on input $n$ \loops
\end{enumerate}
 \end{quote}
For any machine $m$ and any input $n$, exactly one of the above is
true in the standard model and therefore a theorem in \pafor.
	
 \begin{quote} \textbf{Assumption 1}:\emph{\pafor\ is
 semi-decidable.}  That is, we have a machine $G$ which on input
   $\langle p, q\rangle$ outputs \yes\ if $p$ represents a proof in
   \pafor\ of the statement encoded by $q$; otherwise outputs $\no$ or
   \loops.
\end{quote}
If \textbf{Assumption 1} holds, then we can have a machine $H$ which
on input $\langle m,n\rangle$ decides if machine $m$ halts on input
$n$; i.e., $H$ solves the halting problem.  The machine $H$ is
specified below as an algorithm.

\vspace{-0.3in}

\begin{footnotesize}
\begin{algorithm}
\begin{mdframed}[skipabove=0.0in,skipbelow=0.02in,roundcorner=10pt,backgroundcolor=gray!15,
  linewidth=0pt,roundcorner=8pt,fontcolor = black!90,frametitle={
    Algorithm for Machine $H$},frametitlerule=true,frametitlerulecolor=gray!80, frametitlebackgroundcolor=gray!30, frametitlerulewidth=0.5pt]
\caption{Program $H$}
\label{Program H(m,n)}
\SetAlgoLined
\SetKwInOut{Input}{Input}\SetKwInOut{Output}{Output}
\Input{$\langle m,n\rangle$}
\Output{Does $m$ halt on $n$?}
\SetKwBlock{Init}{init}{}
initialization\;
\Init{$q_1$= \textsf{``Arithmetic Statement encoding that $m$ on input $n$ halts with \yes''}\;
$q_2$= \textsf{``Arithmetic Statement encoding that $m$ on input $n$ halts with \no''}\;
$q_3$= \textsf{``Arithmetic Statement encoding that $m$ on input $n$ does not halt or \loops.''}\;
}
\SetKwBlock{Threada}{Thread 1}{}
\SetKwBlock{Threadb}{Thread 2}{}
\SetKwBlock{Threadc}{Thread 3}{}

$H$ is composed of three parallel threads, exactly one of which halts.
If any of the threads halts, $H$ halts.

\Threada{
Do a breadth-first search for a proof $p$ such that 
$G$ on {$\langle p,q_1\rangle$} halts with \yes}

\Threadb{
Do a breadth-first search for a proof $p$ such that 
$G$ on {$\langle p,q_2\rangle$} halts with \yes}

\Threadc{
Do a breadth-first search for a proof $p$ such that 
$G$ on {$\langle p,q_3\rangle$} halts with \yes}
\end{mdframed}
\end{algorithm}	

\vspace{-0.45in}

\begin{algorithm}
\begin{mdframed}[skipabove=0.0in,skipbelow=0.02in,roundcorner=10pt,backgroundcolor=gray!15,
  linewidth=0pt,roundcorner=8pt,fontcolor = black!90,frametitle={
    Algorithm for Breadth-First Search for Finding a Proof for $q$},frametitlerule=true,frametitlerulecolor=gray!80, frametitlebackgroundcolor=gray!30, frametitlerulewidth=0.5pt]
\caption{Breadth-First Search for a Proof}
\SetAlgoLined
Assume we have an lexicographic
ordering of strings $\langle p_0,p_1,...\rangle$.  In the first
iteration we run $G$ on $\langle p_0,q\rangle$ for one step.  In the
next iteration, we run $G$ $\langle p_0,q\rangle$ for one more step
and we also run $G$ on $\langle p_1,q\rangle$ for one step.  We
continue in this fashion until we hit a $p$ such that $G$ on $\langle
p,q\rangle$ stops with \yes.
\end{mdframed}
\end{algorithm}	
\end{footnotesize}

One of the three threads in $H$ will halt.  Therefore $H$ decides the
halting problem.  We have arrived at a contradiction by supposing
\textbf{Assumption 1}, which can be now be discarded, and our
main thesis is established.  \qedsymbol
\end{proof}
\end{small}

\bibliographystyle{agsm}
\bibliography{main72,naveen}

\end{document}